# Deep learning for structural health monitoring: An application to heritage structures

Fabio Carrara, Fabrizio Falchi, Maria Girardi*, Nicola Messina, Cristina Padovani and Daniele Pellegrini

Institute of Information Science and Technologies "A. Faedo", ISTI-CNR, Via Moruzzi 1, 56124 Pisa, Italy

* maria.girardi@isti.cnr.it

**Abstract.** In this work, we employ recent deep learning techniques for time-series forecasting to inspect and detect anomalies in the large dataset recorded during a long-term monitoring campaign conducted on the San Frediano bell tower in Lucca. We frame the problem as an unsupervised anomaly detection task and train a Temporal Fusion Transformer to learn the normal dynamics of the structure. We then detect the anomalies by looking at the differences between the predicted and observed frequencies.

## Introduction

Heritage structures are threatened worldwide by aging, material deterioration, environmental actions, and extreme meteorological events due to climate changes, and therefore they need maintenance and restoration. Structural health monitoring (SHM) provides a relatively inexpensive tool for promptly assessing the structural conditions, planning, and controlling maintenance interventions.

The availability of high-sensitive instrumentation and extensive data sets from long-term monitoring protocols opened new issues about data analysis, particularly regarding the implementation of automatic anomaly detection and early warning procedures. In this data-driven context, exploiting artificial intelligence (AI) represents a significant opportunity. Machine learning (ML) algorithms originated from the broader domain of AI and have recently had a significant diffusion, with applications in many research and industrial areas. ML covers a wide range of algorithms that recognize patterns and build regression models for large multi-source heterogeneous datasets: this approach naturally fits with data collected by SHM systems [1].

Bridges' construction and maintenance represent the classical application fields for automated vibration monitoring. Many papers regarding algorithms for operational modal analysis and damage detection in the domain of ML are available in the literature [2], [3], [4]. The application of ML to the preservation of architectural heritage is relatively recent [5].

In recent years, Deep Learning (DL) techniques have become the state-of-the-art for processing sequential data, like text or audio. The recent astonishing advancements in this field brought important innovations also in the context of time-series processing, obtaining interesting results in time-series forecasting, classification, and anomaly detection [6, 7]. The key to the success of DL in these tasks can be attributed to some neural architectures, such as the Recurrent Neural Networks or the recently introduced Transformer Networks [8]. Despite the fast and pervasive development of DL techniques for time-series processing, relatively few works [9, 10] tried to apply this promising technology to monitor heritage structures and find possible anomalies.

In this paper, we propose to use a recently developed Transformer Network [11] to reveal possible anomalies from the data recorded by the high-sensitive instrumentation installed on heritage structures. We analyze the San Frediano bell tower in Lucca [12], subject to a long-term vibration monitoring campaign from October 2015 to November 2017. The Transformer network is trained to predict the main natural frequencies of the tower starting from some environmental data and the frequencies observed in the very recent past. Anomalous events are then found by comparing how much the actual frequencies deviate from the predicted ones. The data analysis shows promising results, highlighting anomalous events like the Amatrice earthquake that happened on 24 Aug 2016 or the Santa Croce celebrations on 13 Sep 2016.

## Description of the algorithm

The proposed approach belongs to a set of techniques that pertain to the domain of *unsupervised anomaly detection*. Specifically, a network is trained to understand the normality patterns present in some data and then, at inference time, the more significant deviations from the learned normality patterns are considered anomalies. In order to learn the concept of *normality* during the training process, the network is usually trained to predict one or more samples in the future, given observations from the recent past. In this way, the network learns to approximate the normal dynamics of the system under observation. In this work, we use a state-of-the-art deep neural network for learning the structure's dynamics directly from data. This network is called Temporal Fusion Transformer (TFT) [11]. In the following two paragraphs, we discuss how the TFT can predict the structure dynamics while characterizing its uncertainty and how this output can be used for anomaly detection.

*The Temporal Fusion Transformer*. In heritage structure monitoring, data are usually composed of post-processed sensor measurements from the monitored structure, plus some environmental data, like temperature, wind speed, rain, or humidity. The core part of the algorithm is constituted by the network, which predicts the structure's dynamic evolution, given the sensor and environmental data from the recent past. In this study, we use TFT [11], one of the state-of-the-art Transformer Networks for time-series forecasting. Apart from using state-of-the-art attentive modules for processing time-series, the TFT can also estimate the uncertainty of the prediction. This output is fundamental during the anomaly detection phase: we can quickly know whenever the observed actual values fall inside the predicted confidence interval and, if not, easily quantify the deviation from these margins. In the light of this, the system that we develop can be formalized as follows:

$$X_{T+1} = TFT(x_1, x_2, \dots, x_T;\ u_1, u_2, \dots, u_T), \quad (1)$$

where $\boldsymbol{x}_t$ is a vector encoding the values measured from the instrumentation in the structure at time $t$, $\boldsymbol{u}_t$ is a vector encoding the environmental (external) factors like temperature, wind speed, rain, or humidity at time $t$, while the output $\mathbf{X}_{T+1}$ is a random variable encoding the predicted sensor data at time $T+1$. The model characterizes this output random variable by predicting its mean value and 1, 10, 25, 50, 75, 90, 99 percentiles. The training procedure can be treated as a regression problem, where the objective consists of estimating the parameters of the TFT model given some training data. In TFT, the optimization is performed using a quantile loss, which estimates the aleatoric uncertainty of the data. More details in the original TFT paper [11].

*Anomaly Detection*. Once the TFT has been trained on some non-anomalous data, it should have learned the *normal* dynamics of the given system to some extent. Therefore, it should be able to predict the next system state while also quantifying the uncertainty of this prediction. At this point, we can search for anomalous events in a time interval never seen from the network at training time. Given the aleatoric prediction $\mathbf{X}_{T+1}$ and the actual observed state $\boldsymbol{x}_{T+1}$, we can define the state $\boldsymbol{x}_{T+1}$ as *anomalous* if it is an outlier with respect to the estimated distribution of the random variable $\mathbf{X}_{T+1}$. More formally, we can define $\boldsymbol{x}_{T+1}$ as an outlier – or an anomalous sample – by checking if the following condition is NOT satisfied:

$$\pi_{100-p}\,(X_{T+1}) \leq x_{T+1} \leq \pi_p\,(X_{T+1}), \qquad (2)$$

where $\pi_p\,(\mathbf{X})$ represents the $p$-percentile of the random variable $\mathbf{X}$. In our experiments, we use $p = 99$. This procedure is repeated over a sliding window having width $T$ over the whole observation period so that we can predict the presence of anomalies at each timestep $t$ given the sensor and environmental data from the timesteps $\{t\text{-}1,\ t\text{-}2,\ ...,\ t\text{-}T\text{-}1\}$.

**Case study: The San Frediano bell tower**

The San Frediano bell tower in the historic center of Lucca has been subjected to a long-term vibration monitoring campaign from October 2015 to November 2017 [12]. Data from the first year of monitoring (28 Oct 2015 – 13 Aug 2016) are used in the paper to train and validate the TFT model. They were measured by four seismic stations placed alongside the tower's height, each equipped with a tri-axial velocimeter and a 24-bit digitizer of the SARA Electronic Instruments firm. Data were sampled at 100 Hz and continuously acquired over the monitoring period. The training dataset was built by processing data via the covariance-driven Stochastic Subspace Identification technique to extract the first five natural frequencies from the tower's response. We organized data in hourly packages; therefore, the training set consisted of hourly samples of the five frequencies. Together with the frequencies, we considered other environmental variables in the model: air temperature, rainfall and air humidity, wind speed (average and peak values), and wind direction. Furthermore, we conditioned the model on the temporal fingerprints of the observations to better capture periodical patterns at different time scales. Specifically, we conditioned the model on the hour (1-24), the day (1-30), and the month (1-12).

We run the trained model on a time interval not used during the training phase. Specifically, we considered the period from 19 Aug 2016 to 16 Oct 2016, with a window width T = 96 hours (4 days). The results of the analysis are shown in Fig. 1 and Fig.2, reporting the predictions of the model and the observed frequencies. Every prediction is accompanied by the estimated uncertainty, represented by the area between the 1st and 99th percentiles at every prediction timestep. Vertical bars show the time locations when at least one of the frequencies violates the condition expressed in Eq. 2. The color intensity of these lines depends on how much the actual observed value falls outside the estimated uncertainty margins. The final color intensity (yellow low - red high) is obtained by summing the anomaly contributions from each frequency, using the inverse of the frequency as a weight, to attenuate the contribution from higher frequencies. Fig. 1 highlights two major anomalous events, on 24 Aug and 13 Sep 2016. The former event corresponds to the Amatrice earthquake that struck central Italy with a 6.0 magnitude and, despite the distance of 400 km between Amatrice and Lucca, was clearly detected by the sensors installed on the tower and induced vibrations in the same order of magnitude as those caused by the bells' swinging. The latter anomaly corresponds to the Santa Croce celebrations and relates to the presence of many people moving in the town and affecting the tower's vibrations. The algorithm can also detect the vibrations induced by the swinging bells. This anomaly involves the second frequency (related to the swinging direction) and corresponds to the main religious ceremonies held in the Cathedral, particularly on Saturday afternoon and Sunday morning.

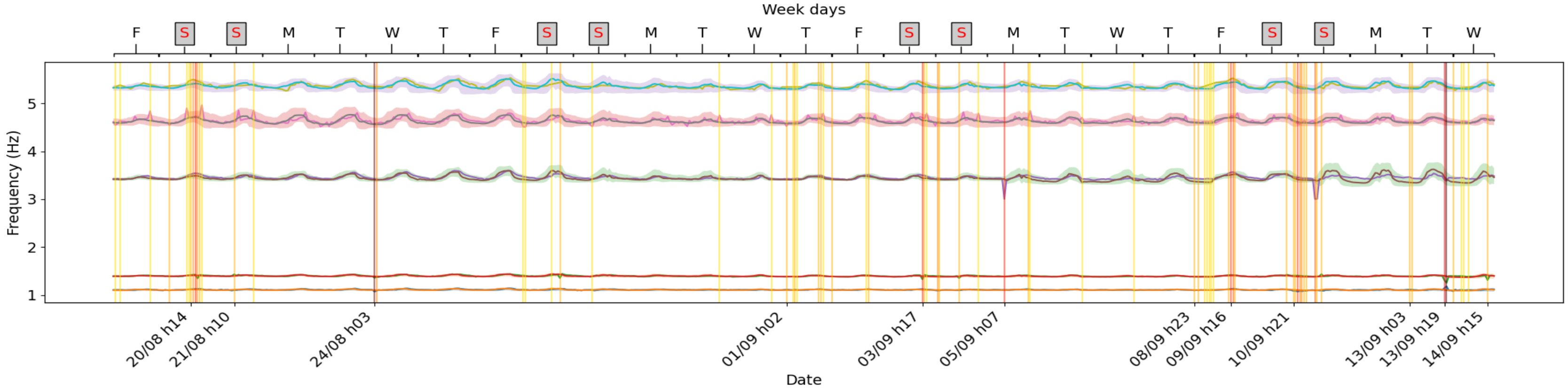

Figure 1: First five predicted and experimental (observed) frequencies, in the period 19 Aug 2016 - 14 Sep 2016.

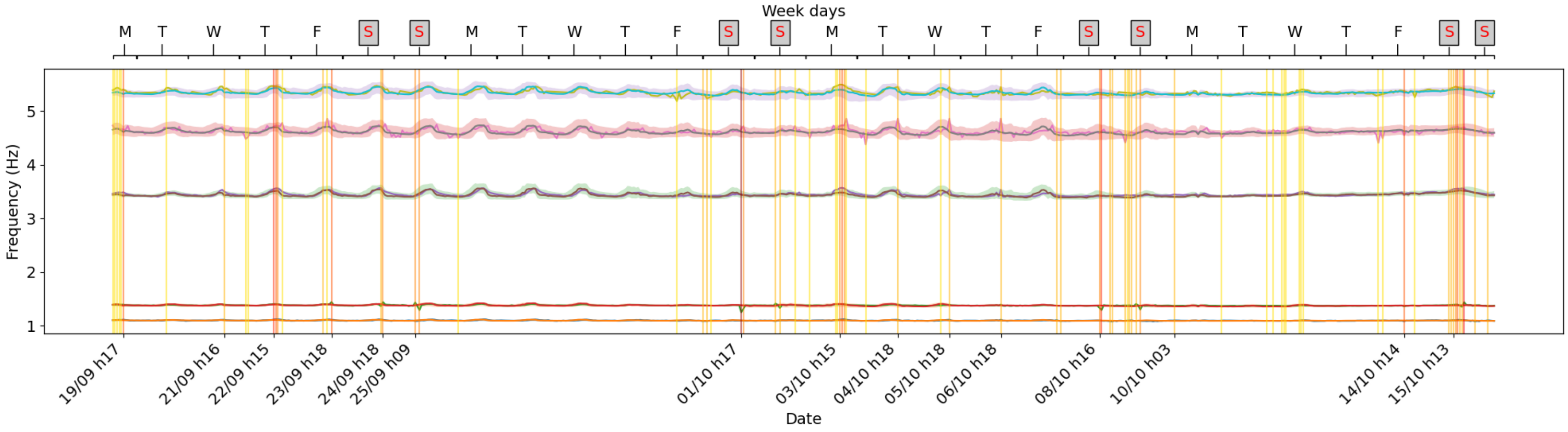

Figure 2: First five predicted and experimental (observed) frequencies, in the period 19 Sep 2016 - 16 Oct 2016.

The anomalous events detected by the TFT procedure are better emphasized in Fig. 3, reporting the observed and predicted values of the first two frequencies from 20 to 26 August 2016, together with the confidence interval evaluated by the algorithm and reported in a dashed line. The anomalies caused by the swinging bells on Saturday (20 Aug, h 17:00) and Sunday (21 Aug, h 10:00 and h 12:00) and by the Amatrice earthquake (24 Aug, h 3:00) are highlighted in the figure; the first two affect mainly the second frequency of the tower, while the effects of the earthquake are evident in the whole dynamic response of the structure.

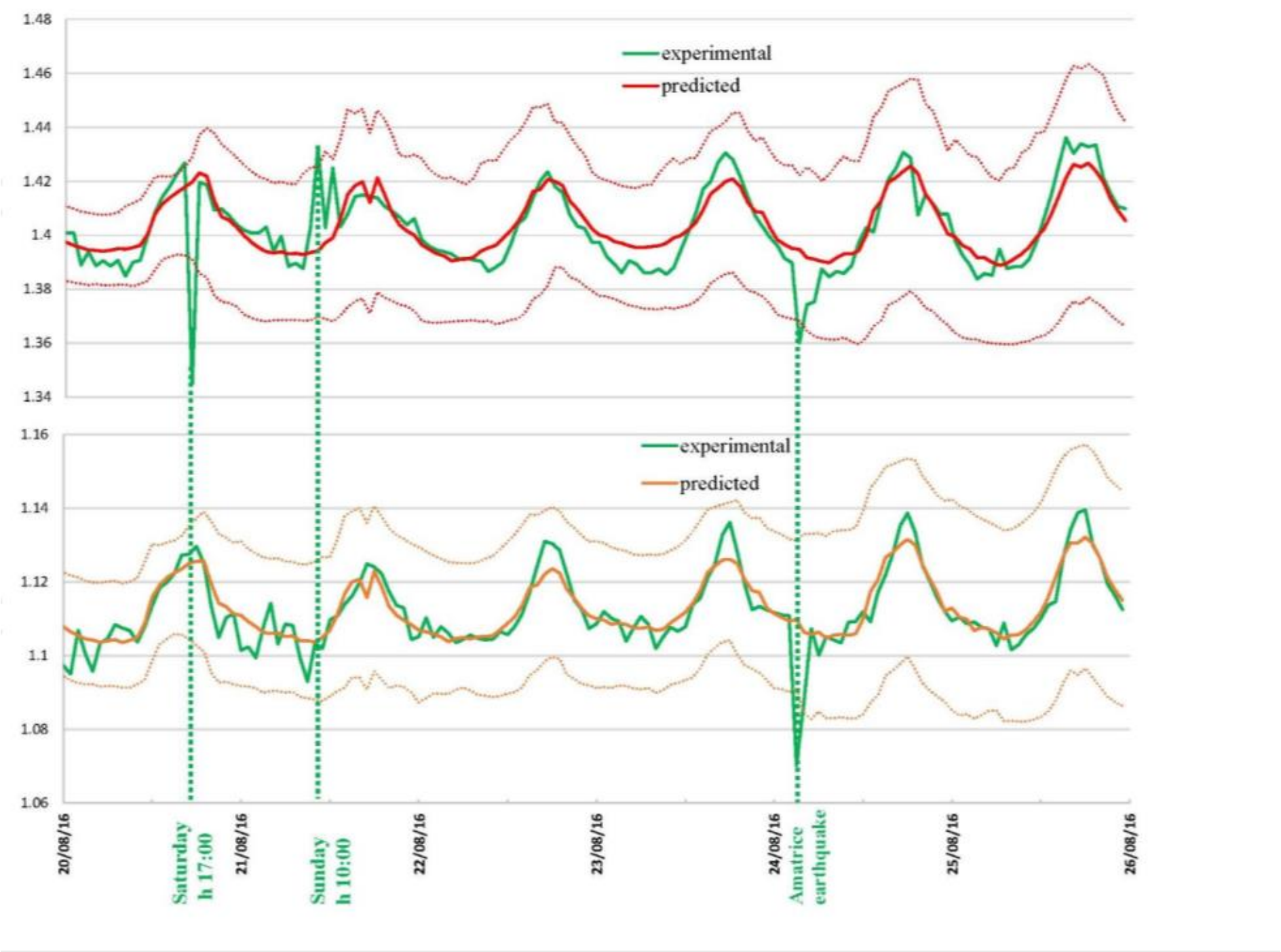

Figure 3: First two predicted and experimental (observed) frequencies, 20-26 Aug 2016.

## Conclusions

In this paper, we explored the potential of state-of-the-art attentive deep neural networks for monitoring heritage structures and spotting possible anomalies. We employed the Temporal Fusion Transformer to learn the normal dynamics of the structure. The tower's dynamic behavior was encoded as its first five natural frequencies, extracted through the covariance-driven Stochastic Subspace Identification technique. The predicted distributions of the frequencies were then used to spot outliers in the test data. This preliminary work showed promising results in the case study of the San Frediano bell tower in Lucca, where we were able to correctly identify some important events, like the Amatrice earthquake and the Santa Croce celebrations.

Further investigations are necessary to confirm the good performance and reliability of the adopted method. A comparison between the anomaly detection techniques used in the literature and the TFT model will be carried out in a future work, where the performance of the Transformer network will be tested on artificial damage scenarios. Furthermore, it would be interesting to characterize the found anomalies by inspecting and clustering the hidden representations learned by the model. This characterization would be helpful, for example, for distinguishing a celebration event happening near the tower from an unexpected structural failure.

**Acknowledgments**
This research has been supported by the Fondazione Cassa di Risparmio di Lucca (SOUL project, 2019-2022), and by the INAROS project (CNR4C program, CUP B53D21008060008). This support is gratefully acknowledged.